\documentstyle[11pt,a4,ere99]{article}
\input{epsf}
\def\be{\begin{equation}}
\def\ee{\end{equation}}
\def\bea{\begin{eqnarray}}
\def\eea{\end{eqnarray}}
\def\b*{\begin{eqnarray*}}
\def\e*{\end{eqnarray*}}

\def\U{\Upsilon}

\def\S{\Sigma}
\def\p{\phi}

\def\Journal#1#2#3#4{{#1} {\bf #2}, #3 (#4)}

\def\NCA{\em Nuovo Cimento}
\def\CR{\em C.R. Acad. Sci. (Paris)}

\def\PRL{\em Phys. Rev. Lett.}

\def\GRG{\em Gen. Rel. Grav.}
\def\CQG{\em Class. Quantum Grav.}

\def\CMP{\em Commun. Math. Phys.}

\begin{document}

\title{Applications of Super-Energy Tensors}

\author{Jos\'e M. M. Senovilla}

\address{Departamento de F\'\i sica Te\'orica, Universidad del
Pa\'\i s Vasco,\\
Apartado 644, 48080 Bilbao, Spain\\E-mail: {\em {\tt wtpmasej@lg.ehu.es}}\\
and\\
Laboratori de F\'\i sica Matem\`atica, Societat Catalana de F\'\i sica, \\
I.E.C., Barcelona, Spain}   




\maketitle\abstracts{
In this contribution I intend to give a summary of the new relevant results
obtained by using the general superenergy tensors. After a quick review of the
definition and properties of these tensors, several of their mathematical and
physical applications are presented. In particular, their interest and
usefulness is mentioned or explicitly analyzed in 1) the study of causal
propagation of general fields; 2) the existence of an infinite number of
conserved quantities in Ricci-flat spacetimes; 3) the different gravitational
theories, such as Einstein's General Relativity or, say, $n=11$ supergravity;
4) the appearance of some scalars possibly related to entropy or quality
factors; 5) the possibility of superenergy exchange between different physical
fields and the appearance of mixed conserved currents.}

\section{Introduction}
In last year ERE-98 meeting, I presented what seemed to be the proper universal
generalization of the traditional Bel and Bel-Robinson super-energy (s-e)
tensors.\cite{S} Since then, a new interest on this subject has developed, as
can be checked in this volume with the contributions by Balfag\'on and
Ja\'en, Bergqvist, Pozo and Parra, and Teyssandier. In this contribution, I
would like to present some of the new relevant applications and results
derived lately by using super-energy tensors, as well as some of the future
developments which are under current consideration. The only pre-requisite to
define the s-e tensors is an $n$-dimensional manifold endowed with a metric of
Lorentzian signature. Thus, all the general results are applicable to most
known theories (including or not the gravitational field), and some
simple examples will be treated here, such as Special and General Relativity,
and $n=11$ supergravity, from where one can also go to, say, type \rm{II}A
string theory.
To that end, I will briefly summarize the definition and properties of the 
general super-energy tensors in the next subsection. For a much more detailed
and comprehensive view, see \cite{S2}.

\subsection{Definition and properties of the s-e tensors}
Let $V_n$ be any differentiable $n$-dimensional manifold endowed with a metric
$g$ of Lorentzian signature (--,+,\dots,+). Indices in $V_n$ run from 0 to
$n-1$ and are denoted by Greek small letters. A useful operation in s-e
studies is the standard Hodge dual, denoted by *, whose definition and
properties can be found at full extent in \cite{S2}. 

The basic idea is to consider any $m$-covariant tensor $t_{\mu_1\dots\mu_m}$
as a so-called {\it r-fold $(n_1,\dots,n_r)$-form}, denoted schematically
by $t_{[n_1],\dots,[n_r]}$, where each of the $[n_{\U}]$ indicates a block
with $n_{\U}$ antisymmetrical indices ($\U=1,\dots,r$).
Several examples of $r$-fold forms are:
$F_{\mu\nu}=F_{[\mu\nu]}$ is a simple 2-form, while $\nabla_{\rho}F_{\mu\nu}$
is a double (1,2)-form; the Riemann tensor is a double {\it symmetrical}
(2,2)-form (the pairs can be interchanged) and the Ricci tensor is a double
symmetrical (1,1)-form; a tensor such as $t_{\mu\nu\rho}=t_{(\mu\nu\rho)}$ is
a triple symmetrical (1,1,1)-form, etcetera.

Then, one can define all possible duals by using the * acting on each
of the $[n_{\U}]$ blocks, obtaining a total of $2^r$ different 
tensors (including $t_{[n_1],\dots,[n_r]}$), each of which is
an $r$-fold form (except when $n_{\U}=n$ for some $\U$, but these special
cases will be treated as self-evident). All these tensors allow to construct
a canonical ``electric-magnetic'' decomposition of $t_{\mu_1\dots\mu_m}$
relative to any observer by means of contraction of each of these duals on all
their $r$ blocks with the timelike unit vector $\vec{u}$ ($u_{\mu}u^{\mu}=-1$)
describing the observer, see \cite{S,S2}: whenever $\vec{u}$ is contracted
with a `starred' block $\stackrel{*}{[n-n_{\U}]}$, we obtain a `magnetic part'
in that block, and an `electric part' otherwise. Thus, the electric-magnetic
parts are (in general) $r$-fold forms which can be denoted by 
\b*
({}^t_{\vec{u}}\underbrace{EE\dots E}_r)_{[n_1-1],[n_2-1],\dots,[n_r-1]},\,\, 
({}^t_{\vec{u}}H\underbrace{E\dots E}_{r-1})_{[n-n_1-1],[n_2-1],\dots,[n_r-1]},\,\,
\dots \\
\dots \, ,\, 
({}^t_{\vec{u}}\underbrace{E\dots E}_{r-1}H)_{[n_1],\dots,[n_{r-1}-1][n-n_r-1]}\, ,\,\,
({}^t_{\vec{u}}HH\underbrace{E\dots E}_{r-2})_{[n-n_1-1],[n-n_2-1],\dots,[n_r-1]},\,\, 
\dots\, ,\\
({}^t_{\vec{u}}\underbrace{E\dots E}_{r-2}HH)_{[n_1-1],\dots,[n-n_{r-1}-1],[n-n_r-1]}\, ,
\, \dots ,
({}^t_{\vec{u}}\underbrace{HH\dots H}_r)_{[n-n_1-1],[n-n_2-1],\dots,[n-n_r-1]},
\e*
where, for instance,
\b*
({}^t_{\vec{u}}\underbrace{EE\dots E}_r)_{\mu_2\dots\mu_{n_1},\nu_2\dots\nu_{n_2},
\dots ,\rho_2\dots\rho_{n_r}}\equiv
\tilde{t}_{\mu_1\mu_2\dots\mu_{n_1},\nu_1\nu_2\dots\nu_{n_2},
\dots ,\rho_1\rho_2\dots\rho_{n_r}}u^{\mu_1}u^{\nu_1}\dots u^{\rho_1}\, ,\\
({}^t_{\vec{u}}H\underbrace{E\dots E}_{r-1})_{\mu_{n_1+2}\dots\mu_{n},\nu_2\dots\nu_{n_2},
\dots ,\rho_2\dots\rho_{n_r}}\!\equiv\! 
\tilde{t}_{\stackrel{*}{\mu_{n_1+1}\mu_{n_1+2}\dots\mu_{n}},
\nu_1\nu_2\dots\nu_{n_2},
\dots ,\rho_1\rho_2\dots\rho_{n_r}}u^{\mu_{n_1+1}}u^{\nu_1}\dots u^{\rho_1}
\e*
and so on. Here, $\tilde{t}_{[n_1],\dots,[n_r]}$ denotes the tensor obtained
from $t_{[n_1],\dots,[n_r]}$ by permutting the indices such that the
first $n_1$ indices are those precisely in the block $[n_1]$, the next
$n_2$ indices are those in the block $[n_2]$, and so on.
There are $2^r$ E-H parts, they are {\it spatial} relative to $\vec{u}$
in the sense that they are orthogonal to $\vec{u}$ in any index,
and all of them determine $t_{\mu_1\dots\mu_m}$ completely.
Besides, $t_{\mu_1\dots\mu_m}$ vanishes iff all its E-H parts do.

Now, the definition of basic s-e tensor for $t_{\mu_1\dots\mu_m}$ is:
\cite{S,S2}
\bea
T_{\lambda_1\mu_1\dots\lambda_r\mu_r}\left\{t\right\}\equiv
\frac{1}{2}\left\{\begin{array}{l} \\ \\ \end{array}\hspace{-4mm}
\left(t_{[n_1],\dots,[n_r]}\times t_{[n_1],\dots,
[n_r]}\right)_{\lambda_1\mu_1\dots\lambda_r\mu_r}+\right. \nonumber \\
+\left(t_{\stackrel{*}{[n-n_1]},\dots,[n_r]}\times
t_{\stackrel{*}{[n-n_1]},\dots,[n_r]}\right)_{\lambda_1\mu_1\dots\lambda_r\mu_r}
+\dots +\nonumber \\
+\dots + \left(t_{[n_1],\dots,\stackrel{*}{[n-n_r]}}\times
t_{[n_1],\dots,\stackrel{*}{[n-n_r]}}\right)_{\lambda_1\mu_1\dots\lambda_r\mu_r}
+ \dots\nonumber \\
+\dots +\left(t_{\stackrel{*}{[n-n_1]},\stackrel{*}{[n-n_2]},\dots,[n_r]}\times
t_{\stackrel{*}{[n-n_1]},\stackrel{*}{[n-n_2]},\dots,[n_r]}
\right)_{\lambda_1\mu_1\dots\lambda_r\mu_r}+\dots +\nonumber\\
\left. +\dots \, \dots +
\left(t_{\stackrel{*}{[n-n_1]},\dots,\stackrel{*}{[n-n_r]}}\times
t_{\stackrel{*}{[n-n_1]},\dots,
\stackrel{*}{[n-n_r]}}\right)_{\lambda_1\mu_1\dots\lambda_r\mu_r}\right\}
\label{set}
\eea
where the $\times$-product is defined for any  $r$-fold
$(n_1,\dots,n_r)$-form by contracting all indices but one of
each block in the product of $\tilde{t}$ with itself, that is to say
\b*
(t\times t)_{\lambda_1\mu_1\dots\lambda_r\mu_r}\equiv
\left(\prod_{\U=1}^{r}\frac{1}{(n_{\U}-1)!}\right)\,
\tilde{t}_{\lambda_1\rho_2\dots\rho_{n_1},\dots ,
\lambda_r\sigma_2\dots\sigma_{n_r}}
\tilde{t}_{\mu_1\hspace{10mm}\dots ,\mu_r}^{\hspace{2mm}\rho_2\dots\rho_{n_1}
,\hspace{5mm}\sigma_2\dots\sigma_{n_r}} \, .
\e*

The s-e tensor (\ref{set}) of $r$-fold forms is therefore a $2r$-covariant
tensor. Notice that any dual of $t_{[n_1],\dots,[n_r]}$ gives rise to the same
basic s-e tensor (\ref{set}). Therefore, one only needs to consider blocks
with at most $n/2$ indices if $n$ is even, or $(n-1)/2$ if $n$ is odd.

The main properties of (\ref{set}) are the following (see \cite{S2} for
explicit proofs):
\begin{enumerate}
\item
Symmetries:
\b*
T_{\lambda_1\mu_1\dots\lambda_r\mu_r}\left\{t\right\}=
T_{(\lambda_1\mu_1)\dots(\lambda_r\mu_r)}\left\{t\right\}
\e*

\item
If the tensor $t_{[n_1],\dots,[n_r]}$ is symmetric in the interchange of the
block $[n_{\U}]$ with the block $[n_{\U'}]$ ($n_{\U}=n_{\U'}$), then
the s-e tensor (\ref{set}) is symmetric in the interchange of the
corresponding $(\lambda_{\U}\mu_{\U})$- and $(\lambda_{\U'}\mu_{\U'})$-pairs.

\item
If $n$ is even, then the s-e tensor (\ref{set}) is traceless in any
$(\lambda_{\U}\mu_{\U})$-pair with $n_{\U}=n/2$.

\item
The {\em super-energy density} of the tensor $t$ relative to the timelike
vector $\vec{u}$ is denoted by $W_t(\vec{u})$ and defined by
\b*
W_t\left(\vec{u}\right)\equiv
T_{\lambda_1\mu_1\dots\lambda_r\mu_r}\{t\}
u^{\lambda_1}u^{\mu_1}\dots u^{\lambda_r}u^{\mu_r} \, .
\e*
Thus, given any unit timelike $\vec{u}$, the s-e density is half the sum of the
positive squares of all the E-H parts of $t$ relative to $\vec{u}$, that is
\b*
W_t\left(\vec{u}\right)=\frac{1}{2}\left(
({}^t_{\vec{u}}\underbrace{EE\dots E}_r)^2+
({}^t_{\vec{u}}H\underbrace{E\dots E}_{r-1})^2+\dots +
({}^t_{\vec{u}}\underbrace{E\dots E}_{r-1}H)^2+\right.\\
\left.+
({}^t_{\vec{u}}HH\underbrace{E\dots E}_{r-2})^2+\dots +
({}^t_{\vec{u}}\underbrace{E\dots E}_{r-2}HH)^2+\dots +
({}^t_{\vec{u}}\underbrace{HH\dots H}_r)^2 \right)
\e*
Then, we have
\b*
\forall \,\,\,\, \mbox{timelike} \,\,\, \vec{u}, \,\,\,\,
W_t(\vec{u})\geq 0, \hspace{2cm}\\
\left\{\exists \vec{u}\hspace{3mm} \mbox{such that} \hspace{2mm}
W_t\left(\vec{u}\right)=0 \right\} \Longleftrightarrow
T_{\lambda_1\mu_1\dots\lambda_r\mu_r}\{t\}=0 \Longleftrightarrow
t_{\mu_1\dots\mu_m}=0 .
\e*

\item
In fact, the s-e density is half the sum of the squares of all the components
of $t$ in any orthonormal basis $\{\vec{e}_{\mu}\}$ with $\vec{e}_0=\vec{u}$:
\b*
W_t(\vec{e}_0)=T_{0\dots 0}\{t\}=\frac{1}{2}\,\,\sum_{\mu_1,\dots,\mu_m=0}^{n-1}
|t_{\mu_1\dots\mu_m}|^2 \, .
\e*

\item
Furthermore \cite{S2}
\b*
W_{t}\left(\vec{u}\right)\propto
\tilde{t}_{\mu_1\dots\mu_{n_1},\dots ,\rho_1\dots\rho_{n_r}}
\tilde{t}_{\nu_1\dots\nu_{n_1},\dots ,\sigma_1\dots\sigma_{n_r}}
h^{\mu_1\nu_1}\dots h^{\mu_{n_1}\nu_{n_1}}\dots
h^{\rho_1\sigma_1}\dots h^{\rho_{n_r}\sigma_{n_r}}
\e*
where $h_{\mu\nu}\left(\vec{u}\right)\equiv g_{\mu\nu}+2u_{\mu}u_{\nu}$.

\item
The {\em super-energy flux vectors} of the tensor $t$ relative to the timelike
vector $\vec{u}$ are denoted by ${}^{\U}\vec{P}_t(\vec{u})$ and defined by
\b*
{}^{\U}P^{\nu}_t\left(\vec{u}\right)\equiv
-\, T_{\lambda_1\mu_1\dots\lambda_{\U-1}\mu_{\U-1}}{}^{\nu}{}_{\mu_{\U}
\dots\lambda_r\mu_r}\{t\}u^{\lambda_1}
u^{\mu_1}\dots u^{\lambda_{\U-1}}u^{\mu_{\U-1}}
u^{\mu_{\U}}\dots u^{\lambda_r}u^{\mu_r} \, .
\e*

The s-e flux vectors can be decomposed with respect to a unit $\vec{u}$ into
their timelike component and the corresponding spatial part as
\b*
{}^{\U}P^{\nu}_t\left(\vec{u}\right)=W_t\left(\vec{u}\right)u^{\nu}+
\left(\delta^{\nu}_{\rho}+u^{\nu}u_{\rho}\right)\,
{}^{\U}P^{\rho}_t\left(\vec{u}\right).
\e*
\item
${}^{\U}\vec{P}_t\left(\vec{u}\right)$ are causal vectors with the same time
orientation than $\vec{u}$.
\item
Some of the above relations for the s-e density and s-e flux vectors are
simple particular cases of a completely general and much more important
property of the basic s-e tensors (\ref{set}), generalizing the
dominant energy condition for energy-momentum tensors \cite{HE} and called the
{\it dominant super-energy property} (DSEP), which reads
\cite{S,S2,BS,Ber3,BerS}
\be
T_{\mu_1\dots\mu_{2r}}\{t\}
k^{\mu_1}_1\dots k^{\mu_{2r}}_{2r} \geq 0 \label{dsep}
\ee
for any future-pointing causal vectors $\vec{k}_1,\dots,\vec{k}_{2r}$. In fact,
the above inequality (\ref{dsep}) is {\it strict} if all the vectors
$\vec{k}_1,\dots,\vec{k}_{2r}$ are timelike. The DSEP (\ref{dsep}) is
equivalent to saying that, in any orthonormal basis $\{\vec{e}_{\nu}\}$,
the `super-energy' relative to $\vec{e}_0$ `dominates' over
all other components of $T_{\mu_1\dots\mu_{2r}}$, that is
\b*
T_{0\dots 0}\geq |T_{\mu_1\dots\mu_{2r}}| \hspace{1cm} \forall
\mu_1,\dots ,\mu_{2r}=0,\dots, n-1 \, .
\e*
\end{enumerate}
It is important to remark that any other linear combination with different
`weights' of the $\times$-products in definition (\ref{set}) will result in the
lose of the DSEP.\cite{S2} This leads to the uniqueness of the completely
symmetric part of (\ref{set}), see \cite{S,S2}.

\subsection{Explicit expressions of s-e tensors in simple cases}

For completeness, let us present the s-e tensors of type (\ref{set})
explicitly for the first numbers $r$. Starting with
a scalar $f$ ($r=0$) we simply have
\b*
T\{f\}=\frac{1}{2}f^2\, .
\e*
For any given simple $p$-form $\S_{\mu_1\dots\mu_p}=\S_{[\mu_1\dots\mu_p]}$
(case $r=1$), the definition (\ref{set}) produces after expanding the duals
\cite{S2}
\be
T_{\lambda\mu}\left\{\S_{[p]}\right\}=\frac{1}{(p-1)!}
\left(\S_{\lambda\rho_2\dots\rho_p}\S_{\mu}{}^{\rho_2\dots\rho_p}-
\frac{1}{2p}g_{\lambda\mu}
\S_{\rho_1\rho_2\dots\rho_p}\S^{\rho_1\rho_2\dots\rho_p}\right).
\label{set1'}
\ee
Consider now any double $(p,q)$-form $K_{[p],[q]}$ (case $r=2$) and take its
corresponding $\tilde{K}_{[p],[q]}$ with ordered indices:
$\tilde{K}_{\mu_1\dots\mu_p,\nu_1\dots\nu_q}=
\tilde{K}_{[\mu_1\dots\mu_p],[\nu_1\dots\nu_q]}$. Its s-e tensor (\ref{set})
reads \cite{S2}
\bea
T_{\alpha\beta\lambda\mu}\left\{K_{[p],[q]}\right\}=
\frac{1}{(p-1)!(q-1)!}\left(
\tilde{K}_{\alpha\rho_2\dots\rho_p,\lambda\sigma_2\dots\sigma_q}
\tilde{K}_{\beta}{}^{\rho_2\dots\rho_p,}{}_{\mu}{}^{\sigma_2\dots\sigma_q}
+\right.\hspace{1cm}\nonumber\\
+\tilde{K}_{\alpha\rho_2\dots\rho_p,\mu\sigma_2\dots\sigma_q}
\tilde{K}_{\beta}{}^{\rho_2\dots\rho_p,}{}_{\lambda}{}^{\sigma_2\dots\sigma_q}
-\frac{1}{p}g_{\alpha\beta}
\tilde{K}_{\rho_1\rho_2\dots\rho_p,\lambda\sigma_2\dots\sigma_q}
\tilde{K}^{\rho_1\rho_2\dots\rho_p,}{}_{\mu}{}^{\sigma_2\dots\sigma_q}
-\hspace{1cm} \label{set11'}\\
\left. -\frac{1}{q}g_{\lambda\mu}
\tilde{K}_{\alpha\rho_2\dots\rho_p,\sigma_1\sigma_2\dots\sigma_q}
\tilde{K}_{\beta}{}^{\rho_2\dots\rho_p,\sigma_1\sigma_2\dots\sigma_q}+
\frac{1}{2pq}g_{\alpha\beta}g_{\lambda\mu}
\tilde{K}_{\rho_1\rho_2\dots\rho_p,\sigma_1\sigma_2\dots\sigma_q}
\tilde{K}^{\rho_1\rho_2\dots\rho_p,\sigma_1\sigma_2\dots\sigma_q}
\right)\nonumber
\eea

One can find similar explicit formulas for the basic s-e tensor
$T_{\alpha\beta\lambda\mu\tau\nu}\left\{A_{[p],[q],[s]}\right\}$ of
general triple $(p,q,s)$-forms,\cite{S2} and so on.
It is noteworthy that expressions (\ref{set1'}) and (\ref{set11'})
do not depend on the dimension $n$ explicitly. This is in
fact a general property, so that a formula for the s-e tensor (\ref{set})
can be given, in general, without any explicit dependence on $n$. However,
for $n=4$, the above s-e tensors adopt a very simple and illuminating explicit
expression by using spinors, see \cite{Ber3}. Another simple expression in
general $n$ can be produced by using Clifford algebra techniques, see
Pozo and Parra's contribution.\cite{PP}

\section{Applications of super-energy tensors}
One of the main applications of s-e tensors is to the study of the causal
propagation of general fields. In this study, the use of the DSEP is very
helpful, and thus very simple conditions on the divergence of the s-e tensor
of a field can be found for the field to propagate causally. We refer the
reader to \cite{BerS,BS} and to Bergqvist's contribution to this volume.
Other important applications concern the existence of Bel-Robinson-type tensors
for physical fields other than gravity, to the definition of scalars of mathematical interest, and to the existence of new
conservation laws. These are considered briefly in what follows.

\subsection{The massive scalar field: (super)$^k$-energy tensors}
Consider now a scalar field $\phi$ with mass $m$ (the massless case is included
by setting $m=0$) satisfying the Klein-Gordon equation
\be
\nabla_{\rho}\nabla^{\rho}\p=m^2\p \, .\label{KGm}
\ee
Its energy-momentum tensor reads
\be
T_{\lambda\mu}=\nabla_{\lambda}\phi\nabla_{\mu}\phi
-\frac{1}{2}g_{\lambda\mu}\nabla_{\rho}\phi\nabla^{\rho}\phi 
-\frac{1}{2}g_{\lambda\mu}m^2\p^2 \label{emsm}
\ee
which is symmetric and identically divergence-free when (\ref{KGm}) holds.
Actually, it is remarkable that (\ref{emsm}) can be written as
\b*
T_{\lambda\mu}=T_{\lambda\mu}\left\{\nabla_{[1]}\p\right\}+
T\left\{m \p\right\}(-g_{\lambda\mu}).
\e*
This procedure is systematic \cite{S2} and one can produce tensors with a
higher number of indices by constructing the s-e tensors of type (\ref{set})
for the higher-order derivatives of $\p$. For instance, one can consider the
next step by using $\nabla_{[1]}\nabla_{[1]}\p$ as starting object and adding
the corresponding mass terms: \cite{S2}
\bea
{\cal S}_{\alpha\beta\lambda\mu}\equiv
T_{\alpha\beta\lambda\mu}\left\{\nabla_{[1]}\nabla_{[1]}\p\right\}+
T_{\alpha\beta}\left\{\nabla_{[1]}\left(m\p\right)\right\}(-g_{\lambda\mu})+
T_{\lambda\mu}\left\{m\left(\nabla_{[1]}\p\right)\right\} (-g_{\alpha\beta})+
\nonumber\\
+T\left\{m\, m\, \p \right\}(-g_{\alpha\beta})(-g_{\lambda\mu})=
2\nabla_{\alpha}\nabla_{(\lambda}\phi \nabla_{\mu)}\nabla_{\beta}\phi
-g_{\alpha\beta}\left(\nabla_{\lambda}\nabla^{\rho}\phi
\nabla_{\mu}\nabla_{\rho}\phi+m^2\nabla_{\lambda}\phi\nabla_{\mu}\phi \right)
\nonumber \\
- g_{\lambda\mu}\left(\nabla_{\alpha}\nabla^{\rho}\phi
\nabla_{\beta}\nabla_{\rho}\phi
+m^2\nabla_{\alpha}\phi\nabla_{\beta}\phi\right)
+\frac{1}{2} g_{\alpha\beta}g_{\lambda\mu}\left(
\nabla_{\sigma}\nabla_{\rho}\phi\nabla^{\sigma}\nabla^{\rho}\phi +
2m^2\nabla_{\rho}\phi\nabla^{\rho}\phi +m^4\phi^2\right) \label{sesm}
\eea
which satisfies
\be
{\cal S}_{\alpha\beta\lambda\mu}={\cal S}_{(\alpha\beta)(\lambda\mu)}=
{\cal S}_{\lambda\mu\alpha\beta}.\label{sesmsym}
\ee
The tensor (\ref{sesm}) has been previously considered by Bel (unpublished) and
Teyssandier \cite{Tey} for the case of Special Relativity ($n=4$). The 
resemblance of this tensor with the Bel tensor (see \cite{B4,Beltesis,BS2} and
the next subsection) has led to consider it as a {\it super-energy tensor} of
the scalar field, describing similar properties as the Bel-Robinson tensor does
for the gravitational field. In fact, this correspondence can be sustained
on mathematical and physical grounds \cite{S2}, and moreover it can be carried
out further to a cascade of infinite {\it (super)$^k$-energy} tensors, one for
each natural number $k$, by considering the $(k+1)^{th}$ covariant derivative
of the scalar field as starting object.\cite{S2} An important result is that
the (super)$^k$-energy (tensor) of $\p$ vanishes at a point $x\in V_n$ if and
only if $\p$ and all its derivatives up to the $(k+1)^{th}$ order are zero at
$x$. In particular, {\em all} these (super)$^k$-energy tensors
vanish in a domain $D\subseteq V_n$ if $\p$ vanishes in $D$.

Even more interesting results can be drawn from the study of the divergence of
(\ref{sesm}). A straightforward calculation leads to\cite{S2}
\bea
\nabla_{\alpha}{\cal S}^{\alpha}{}_{\beta\lambda\mu}=
2\nabla_{\beta}\nabla_{(\lambda}\phi
R_{\mu)\rho}\nabla^{\rho}\phi -g_{\lambda\mu} R^{\sigma\rho}
\nabla_{\beta}\nabla_{\rho}\phi\nabla_{\sigma}\phi -\nonumber\\
-\nabla_{\sigma}\phi \left(2\nabla^{\rho}\nabla_{(\lambda}\phi \,
R^{\sigma}_{\mu)\rho\beta} +g_{\lambda\mu}
R^{\sigma}_{\rho\beta\tau}\nabla^{\rho}\nabla^{\tau}\phi\right)
\label{divsesm}
\eea
so that it is immediate to see that ${\cal S}$ is divergence-free in {\it flat
spacetimes}, which leads to the existence of conserved quantities for the
scalar field in the absence of curvature. For a study of these in Special
Relativity see \cite{Tey}. In fact, the above divergence-free result holds
for all (super)$^k$-energy tensors, leading to an infinite number of conserved
quantities in flat spacetimes.\cite{S2,S3}

\subsection{The gravitational field: generalized Bel and Bel-Robinson tensors}
Assume that the gravitational field is described by the Riemman tensor
$R_{\alpha\beta\lambda\mu}$ of the spacetime. As this is a double symmetric
(2,2)-form, the basic s-e tensor for the gravitational field is simply given
by the appropriate restriction of (\ref{set11'})
\bea
B_{\alpha\beta\lambda\mu}\equiv 
T_{\alpha\beta\lambda\mu}\left\{R_{[2],[2]}\right\}=
R_{\alpha\rho,\lambda\sigma}
R_{\beta}{}^{\rho,}{}_{\mu}{}^{\sigma}
+R_{\alpha\rho,\mu\sigma}
R_{\beta}{}^{\rho,}{}_{\lambda}{}^{\sigma}-\nonumber\\
-\frac{1}{2}g_{\alpha\beta}
R_{\rho\tau,\lambda\sigma}R^{\rho\tau,}{}_{\mu}{}^{\sigma}
-\frac{1}{2}g_{\lambda\mu}
R_{\alpha\rho,\sigma\tau}R_{\beta}{}^{\rho,\sigma\tau}+
\frac{1}{8}g_{\alpha\beta}g_{\lambda\mu}
R_{\rho\tau,\sigma\nu}
R^{\rho\tau,\sigma\nu} \label{Bel}
\eea
which is a straightforward generalization of the original definition
\cite{B4,Beltesis} of the so-called Bel tensor (see \cite{C}
for $n=4$, and also \cite{Gal}). The properties of (\ref{Bel}) are
\be
B_{\alpha\beta\lambda\mu}=B_{(\alpha\beta)(\lambda\mu)}=
B_{\lambda\mu\alpha\beta}.\label{belsym}
\ee
and also
\be
\nabla_{\alpha}B^{\alpha\beta\lambda\mu}=
R^{\beta\hspace{1mm}\lambda}_{\hspace{1mm}\rho\hspace{2mm}\sigma}
J^{\mu\sigma\rho}+R^{\beta\hspace{1mm}\mu}_{\hspace{1mm}\rho\hspace{2mm}\sigma}
J^{\lambda\sigma\rho}-\frac{1}{2}g^{\lambda\mu}
R^{\beta}_{\hspace{1mm}\rho\sigma\gamma}J^{\sigma\gamma\rho}\label{divbel}
\ee
where $J_{\lambda\mu\beta}=-J_{\mu\lambda\beta}\equiv
\nabla_{\lambda}R_{\mu\beta}-\nabla_{\mu}R_{\lambda\beta}$. Thus, $B$ is
{\it divergence-free} when the `current' of matter
$J_{\lambda\mu\beta}$ vanishes. 

One can also construct
the basic s-e tensor for the Weyl tensor, which generalizes the classical
Bel-Robinson tensor \cite{B1,PR,BS2} constructed in General Relativity,
giving
\bea
{\cal T}_{\alpha\beta\lambda\mu}\equiv 
T_{\alpha\beta\lambda\mu}\left\{C_{[2],[2]}\right\}=
C_{\alpha\rho,\lambda\sigma}
C_{\beta}{}^{\rho,}{}_{\mu}{}^{\sigma}
+C_{\alpha\rho,\mu\sigma}
C_{\beta}{}^{\rho,}{}_{\lambda}{}^{\sigma}-\nonumber\\
-\frac{1}{2}g_{\alpha\beta}
C_{\rho\tau,\lambda\sigma}C^{\rho\tau,}{}_{\mu}{}^{\sigma}
-\frac{1}{2}g_{\lambda\mu}
C_{\alpha\rho,\sigma\tau}C_{\beta}{}^{\rho,\sigma\tau}+
\frac{1}{8}g_{\alpha\beta}g_{\lambda\mu}
C_{\rho\tau,\sigma\nu}
C^{\rho\tau,\sigma\nu} \label{BR}
\eea
from where one deduces the same symmetry properties as in (\ref{belsym}). One
is used to the properties that the Bel-Robinson tensor is completely symmetric
and traceless, but these depend on the dimension $n$ of the spacetime. In fact,
it can be proven\cite{S2} that (\ref{BR}) is completely symmetric if and only
if $n=4,5$, and traceless iff $n=4$. Nevertheless, the divergence of
(\ref{BR}) vanishes in general $n$-dimensional Einstein spaces.

The difference between the Riemann and Weyl tensors is described by the Ricci
tensor, which is usually related to the matter contents of the spacetime (as
in General Relativity through Einstein's equations). Thus, one can define a
{\it pure matter} gravitational s-e tensor by means of \cite{S2}
\b*
{\cal M}_{\alpha\beta\lambda\mu}\equiv
T_{\alpha\beta\lambda\mu}\left\{R_{[2],[2]}-C_{[2],[2]}\right\},
\hspace{1cm} {\cal M}_{\alpha\beta\lambda\mu}=
{\cal M}_{(\alpha\beta)(\lambda\mu)}=
{\cal M}_{\lambda\mu\alpha\beta}
\e*
which has the interesting property of vanishing iff the spacetime is Ricci
flat. Taking into account the classical decomposition of the Riemann tensor
into the Weyl tensor and the Ricci terms, one can easily prove
\b*
B_{\alpha\beta\lambda\mu}={\cal T}_{\alpha\beta\lambda\mu}+
{\cal M}_{\alpha\beta\lambda\mu}+{\cal Q}_{\alpha\beta\lambda\mu}
\e*
where ${\cal Q}_{\alpha\beta\lambda\mu}$ contains the coupled terms and
satisfies (for the explicit expression see \cite{S2})
\be
{\cal Q}_{\alpha\beta\lambda\mu}={\cal Q}_{(\alpha\beta)(\lambda\mu)}=
{\cal Q}_{\lambda\mu\alpha\beta},\hspace{5mm}
{\cal Q}_{\alpha(\beta\lambda\mu)}=0.\label{Q}
\ee
The last property in (\ref{Q}) allows to prove that,
in fact, the tensor ${\cal Q}$ does {\it not} contribute to the generalized
Bel s-e flux vector, so that (using obvious notation)
\b*
\vec{P}_{B}\left(\vec{u}\right)=
\vec{P}_{{\cal T}}\left(\vec{u}\right)+\vec{P}_{{\cal M}}\left(\vec{u}\right)
\hspace{1cm}
W_{B}\left(\vec{u}\right)=
W_{{\cal T}}\left(\vec{u}\right)+W_{{\cal M}}\left(\vec{u}\right),
\e*
from where one can also easily get a characterization of Ricci-flat spacetimes
\b*
\left\{\exists \vec{u}\hspace{3mm} \mbox{such that} \hspace{2mm}
W_{{\cal M}}\left(\vec{u}\right)=0 \right\} \Longleftrightarrow
{\cal M}_{\alpha\beta\lambda\mu}=0 \Longleftrightarrow \\
\Longleftrightarrow R_{\alpha\beta,\lambda\mu}=C_{\alpha\beta,\lambda\mu} 
\Longleftrightarrow R_{\mu\nu}=0 \Longleftrightarrow 
B_{\alpha\beta\lambda\mu}={\cal T}_{\alpha\beta\lambda\mu}\, .
\e*

The properties of the different gravitational s-e densities allow to compare
the relative strength of the Riemman, Weyl and Ricci tensors in a given
spacetime. For instance, the three positive scalars relative to an observer
$\vec{u}$
\b*
q_1\equiv \frac{W_{{\cal M}}\left(\vec{u}\right)}{W_{B}\left(\vec{u}\right)},
\hspace{1cm}
q_2\equiv
\frac{W_{{\cal T}}\left(\vec{u}\right)}{W_{{\cal M}}\left(\vec{u}\right)},
\hspace{1cm}
q_3\equiv \frac{W_{{\cal T}}\left(\vec{u}\right)}{W_{B}\left(\vec{u}\right)}
=q_1q_2
\e*
are not independent in general, and they satisfy\cite{S2}
\b*
0\leq q_1\leq 1, \hspace{1cm} 0\leq q_2\leq \infty , \hspace{1cm}
0\leq q_3\leq 1 .
\e*
Thus, $q_2$ can be used to analyze the so-called Penrose's Weyl tensor
hypothesis \cite{P6,W3} and similar conjectures concerning
the entropy of the gravitational field \cite{Bon}. On the other hand,
$q_1$ may be a `quality factor' for approximate solutions of some field
equations following the ideas in \cite{calidad}.
Finally, the scalar $q_3$ vanishes if and only if 
the metric is conformally flat, so that $q_3$ measures the departure from
this condition somehow.

Similarly to the case of the scalar field, one can also define the
(super)$^k$-energy tensors for the gravitational field by using the
$(k-1)^{th}$ covariant derivative of the Riemann tensor as starting object.
These tensors are related for each $k$ with the corresponding $k$-level of the
scalar and other fields \cite{S2} and may have some relevance at points of
$V_n$ where the Riemann tensor vanishes but such that some of its derivatives
do not (so that every neighbourhood of the point has non-zero curvature).
Moreover, the gravitational (super)$^k$-energy (tensor) vanishes in a domain
$D\subseteq V_n$ iff the $(k-1)^{th}$ covariant derivative of the Riemann
tensor is zero in $D$. In particular, {\em all} gravitational
(super)$^k$-energy tensors vanish in flat regions of $(V_n,g)$.

All this provides an intrinsic
characterization of the $(V_n,g)$ of constant curvature as those with
vanishing gravitational (super)$^2$-energy (tensor), so that de Sitter and
anti-de Sitter spacetimes can be defined as those spacetimes having identically
vanishing (super)$^2$-energy but non-zero super-energy.

\subsection{n=11 Supergravity}
As a simple example of the potentialities of the s-e construction, let us
apply it to the now fashionable 11-dimensional Supergravity (11-SUGRA),
considering only the bosonic sector for simplicity. As is known, see e.g.
\cite{Mar}, this includes only two fields, the gravitational one described by
the Riemann tensor as usual, and a gauge field
$F_{\alpha\beta\gamma\delta}=F_{[\alpha\beta\gamma\delta]}$ which is a 
simple 4-form. Therefore, one can define the s-e tensor for 11-SUGRA as
the generalized Bel tensor (\ref{Bel}) for $n=11$ combined with the s-e
tensor constructed for the double (1,4)-form $\nabla_{[1]}F_{[4]}$, which 
using (\ref{set11'}) reads
\b*
T_{\alpha\beta\lambda\mu}\left\{\nabla_{[1]}F_{[4]}\right\}=\frac{1}{3!}
\left(\nabla_{\alpha}F_{\lambda\rho\sigma\tau}
\nabla_{\beta}F_{\mu}{}^{\rho\sigma\tau}+
\nabla_{\alpha}F_{\mu\rho\sigma\tau}
\nabla_{\beta}F_{\lambda}{}^{\rho\sigma\tau}-\frac{}{}\right. \\
-\left. g_{\alpha\beta}
\nabla_{\nu}F_{\lambda\rho\sigma\tau}
\nabla^{\nu}F_{\mu}{}^{\rho\sigma\tau}-\frac{1}{4}g_{\lambda\mu}
\nabla_{\alpha}F_{\nu\rho\sigma\tau}
\nabla_{\beta}F^{\nu\rho\sigma\tau}+\frac{1}{8}g_{\alpha\beta}g_{\lambda\mu}
\nabla_{\zeta}F_{\nu\rho\sigma\tau}
\nabla^{\zeta}F^{\nu\rho\sigma\tau}\right).
\e*
Thus, the total s-e tensor for 11-SUGRA should read as
\b*
T_{\alpha\beta\lambda\mu}\left\{\mbox{11-SUGRA}\right\}\equiv
B_{\alpha\beta\lambda\mu}+\kappa \,
T_{\alpha\beta\lambda\mu}\left\{\nabla_{[1]}F_{[4]}\right\}
\e*
where $\kappa >0$ is an available positive constant, possibly related to the
coupling constant $k_{11}$, with the appropriate physical units. 

From this expression one can also go to type \rm{II}A String theory by
projecting \cite{Mar}, or alteratively one can construct the s-e tensor for
\rm{II}A String by starting there. In fact, one should compare the
different s-e tensors obtained by these two methods. The relevance of these
s-e tensors for the string and other higher dimensional theories is under
current investigation. Compare with \cite{DK,DKS,Des,Des2,DS}.

\section{Conserved currents for the Einstein-Klein-Gordon case}
Perhaps the most important physical application of the s-e tensors is the
possibility that arises of exchange of super-energy-momentum quantities between
different physical fields. Herein, we will only consider the case of a
scalar field minimally coupled to gravity, that is, the case when the
Einstein-Klein-Gordon equations hold. These equations can be written in
general dimension $n$ as
\be
R_{\mu\nu}=\nabla_{\mu}\phi\nabla_{\nu}\phi +
\frac{1}{n-2}m^2\phi^2g_{\mu\nu}
\label{ric}
\ee
from where one can deduce the Klein-Gordon equation (\ref{KGm}).

Remember that the situation is as follows: the Bel tensor is divergence-free
in Ricci-flat spacetimes (that is, if there is no matter and the Einstein
equations hold), and the s-e tensor (\ref{sesm}) is divergence-free in
the absence of curvature, which can be interpreted as absence of gravitational
field. These divergence-free properties lead to conserved currents (i.e., 
divergence-free vector fields) whenever there are symmetries in the spacetime,
see \cite{S2} for a lengthy discussion. Thus, the natural question arises
of whether or not one can combine the two s-e tensors to produce a conserved
current in the mixed case: when there are both a scalar field and the
curvature that it generates. And the answer is, in general, affirmative
\cite{S3,S2}.

In order to prove it, let us assume that the spacetime has a Killing vector
${\vec \xi}$. Then, it is known \cite{Pa,Sh} that
\bea
\xi^{\mu}\nabla_{\mu}\phi =0, \hspace{1cm} \mbox{(if $m\neq 0$)},\nonumber\\
\xi^{\beta} \nabla^{\rho}\p\, \nabla_{\beta}\nabla_{\rho}\p  = 0.
\hspace{1cm} \label{xpp}
\eea
If the scalar field is massless, then in fact one has
$\xi^{\mu}\nabla_{\mu}\phi =$const.\, see \cite{Pa,Sh}. In any case,
(\ref{xpp}) always holds. Using (\ref{ric}) one deduces
\b*
J_{\lambda\mu\beta}=\nabla_{\beta}\nabla_{\lambda}\p\,
\nabla_{\mu}\p-\nabla_{\beta}\nabla_{\mu}\p\, \nabla_{\lambda}\p+
\frac{2}{n-2} m^2\p
\left(g_{\beta\mu}\nabla_{\lambda}\p -g_{\beta\lambda}\nabla_{\mu}\p\right)
\e*
so that contracting (\ref{divbel}) and (\ref{divsesm}) with three copies of
${\vec \xi}$ and using the above one gets
\b*
\xi^{\beta}\xi^{\lambda}\xi^{\mu}\,
\nabla_{\alpha}B^{\alpha}{}_{\beta\lambda\mu} 
=\nabla_{\sigma}\phi 
\left(2\nabla_{\rho}\nabla_{(\lambda}\phi
R^{\sigma}{}_{\mu}{}^{\rho}{}_{\beta)}
+g_{(\lambda\mu}R^{\sigma\rho}{}_{\beta)}{}^{\tau}
\nabla_{\rho}\nabla_{\tau}\phi\right)\xi^{\beta}\xi^{\lambda}\xi^{\mu},\\
\xi^{\beta}\xi^{\lambda}\xi^{\mu}\,
\nabla_{\alpha}{\cal S}^{\alpha}{}_{\beta\lambda\mu} 
=-\nabla_{\sigma}\phi 
\left(2\nabla_{\rho}\nabla_{(\lambda}\phi
R^{\sigma}{}_{\mu}{}^{\rho}{}_{\beta)}
+g_{(\lambda\mu}R^{\sigma\rho}{}_{\beta)}{}^{\tau}
\nabla_{\rho}\nabla_{\tau}\phi\right)\xi^{\beta}\xi^{\lambda}\xi^{\mu}
\e*
and, in general, neither of these quantities is zero. However, as is obvious
\b*
\xi^{\beta}\xi^{\lambda}\xi^{\mu}\nabla_{\alpha}\left(
B^{\alpha}{}_{\beta\lambda\mu} 
+{\cal S}^{\alpha}{}_{\beta\lambda\mu}\right)=0
\e*
so that, using (\ref{sesmsym}), (\ref{belsym}) and
$\nabla_{(\mu}\xi_{\nu)}=0$ one can finally write \cite{S3,S2}
\b*
\fbox{$\nabla_{\alpha}j^{\alpha}=0$} \hspace{1cm}
j^{\alpha}\equiv \left(B^{\alpha\beta\lambda\mu}+
{\cal S}^{\alpha\beta\lambda\mu}\right)\xi_{\beta}\xi_{\lambda}\xi_{\mu}\, .
\e*
Notice that only the completely symmetric parts of $B$ and ${\cal S}$ are
relevant here. The importance of this result is that
provides {\it conserved s-e quantities} (via a typical
integration of $j^{\alpha}$ and Gauss' theorem) {\it proving the exchange
of s-e properties} between the gravitational and scalar fields, because neither
$B^{\alpha\beta\lambda\mu}\xi_{\beta}\xi_{\lambda}\xi_{\mu}$ nor
${\cal S}^{\alpha\beta\lambda\mu}\xi_{\beta}\xi_{\lambda}\xi_{\mu}$ are
divergence-free separately in general.

Actually, one can generalize the above result by contracting with
different Killing vectors (if they exist).\cite{S2} Furthermore, the
interchange of s-e quantities between the electromagnetic
and the gravitational field can also be proven by analyzing the propagation
of discontinuities along null hypersurfaces \cite{S,S2,S3}. In fact, this 
last analysis shows that the different (super)$^k$-energy levels give also
raise to conserved mixed quantities. 

The expressions of the above divergence-free currents for some explicit
spacetimes and the physical properties of the mixed conserved quantities
thus generated are under current study. All in all, it seems that the s-e
tensors defined last year in \cite{S} are leading to interesting results and
applications in several disconnected directions, and their future looks very
promising.

\section*{Acknowledgments}
Financial support from the Basque Country University under project
number UPV172.310-G02/99 is acknowledged.

\end{document}